\newcounter{myctr}
\def\myitem{\refstepcounter{myctr}\bibfont\noindent\ifnum\themyctr>9\else\phantom{0}\fi\hangindent17pt\themyctr.\enskip}

\documentclass{ws-ijqi}
 
\begin{document}

\markboth{F.~Buscemi, P.~Bordone and A.~Bertoni}
{Entanglement creation in semiconductor quantum dot charge qubit}

%%%%%%%%%%%%%%%%%%%%% Publisher's Area please ignore %%%%%%%%%%%%%%
%\catchline{}{}{}{}{}
%%%%%%%%%%%%%%%%%%%%%%%%%%%%%%%%%%%%%%%%%%%%%%%%%%%%%%%%%%%%%%%%%%%

\title{ENTANGLEMENT CREATION IN SEMICONDUCTOR \\QUANTUM DOT CHARGE QUBIT }

\author{FABRIZIO BUSCEMI}

\address{ Advanced Research Center on Electronic Systems  and Department of Electronics, \\University of Bologna, Via Toffano 2/2, 40125 Bologna, Italy \\
  fabrizio.buscemi@unimore.it}

\author{PAOLO BORDONE}

\address{Department of Physics, University of Modena and Reggio
  Emilia, \\Via Campi 213/A, 41125 Modena, Italy}

\author{ANDREA BERTONI} \address{Center S3, CNR-Institute of
  Nanosciences, Via Campi 213/A, 41125 Modena, Italy}

\maketitle

% \begin{history}
%   \received{Day Month Year} \revised{Day Month Year}
% %   \accepted{Day Month Year} \comby{(xxxxxxxxxx)}
% \end{history}

\begin{abstract}
We study theoretically  the appearance
of quantum correlations   in two- and three-electron
scattering in single and double dots.
The key role played by transport resonances into entanglement
formation between the single-particle  states is shown.
Both reflected and transmitted components  of  the 
scattered particle wavefunction
are used to evaluate the quantum correlations between 
the incident carrier and the bound particle(s) in the dots.
Our investigation  provides  a  guideline
for the analysis of decoherence effects due to the Coulomb scattering
in semiconductor quantum dots structures.
\end{abstract}


\section{Introduction}
Recent advances in nanofabrication technology  permit to design
atom-like structures, such as quantum dots (QDs) and double quantum dot
(DQDs), where electron states can  be controlled and
manipulated.\cite{Haya,Shink}  Specifically, the charge states of QDs and DQDs have
been shown to be excellent prototype blocks for the implementation of
quantum-computing solid-state devices.\cite{Div,Div2} Indeed, they are
scalable to large system and compatible with current microelectronics
technology.

A number of works address the entanglement production, control, and
detection in single and double dots nanostructures on the basis of
different physical mechanisms.\cite{Oli,raa,Lopez,Buscemi,busce2} In all of them the most serious
obstacle into generating entanglement and therefore performant quantum
circuits is  the interaction between the quantum system and its
environment, resulting in the system decoherence.  For charge qubits in
QD and DQD the channels of decoherence are mainly two: the
electron-phonon coupling and the electron-electron Coulomb interaction.\cite{Stavrou} The former,
stemming from the interaction with the  crystal lattice, is
widely present in all nanostructures even if its effects can be small
at very low temperatures. The latter is due the coupling of the bound
electrons in DQ and DQD to other  carriers in the system
and is therefore extrinsic.

In this work we  analyze the entanglement formation due to
two- and three-particle scattering events in QD and DQD structures,
respectively. To this aim, we consider collisions between one or two
electrons localized in a single or double dot and another electron
entering the structure from a lead, and solve the corresponding Schr\"odinger equation of  the two- or three-electron wavefunction in the real-space representation.
In our numerical  approach the scattering states are computed
with the  temperature of the system  set  equal to zero
in order to realistically neglect the  electron-phonon
interaction. 
Specifically, we obtain the
transmission (TC) and reflection coefficient (RC) of the electron crossing the
nanostructure as a function of its  initial kinetic energy and of the orbital state of the bound electron(s).\cite{Buscemi,busce2,Bertoni}
 We
stress that the estimation of the amount of  quantum  correlations
is a measure of the loss of coherence undergone by the charge qubits
of the DQ and DQD when another carrier interacts with them.

\section{The Physical  Models }
We compute the amount of quantum correlations between the  scattered and the bound carriers interacting through  Coulomb repulsion in two different
systems, namely a single QD and a double QD, as described in the following.
\subsection{Single-dot scattering}
\begin{figure}[htpb]
  \begin{center}
    \includegraphics*[width=0.5\linewidth]{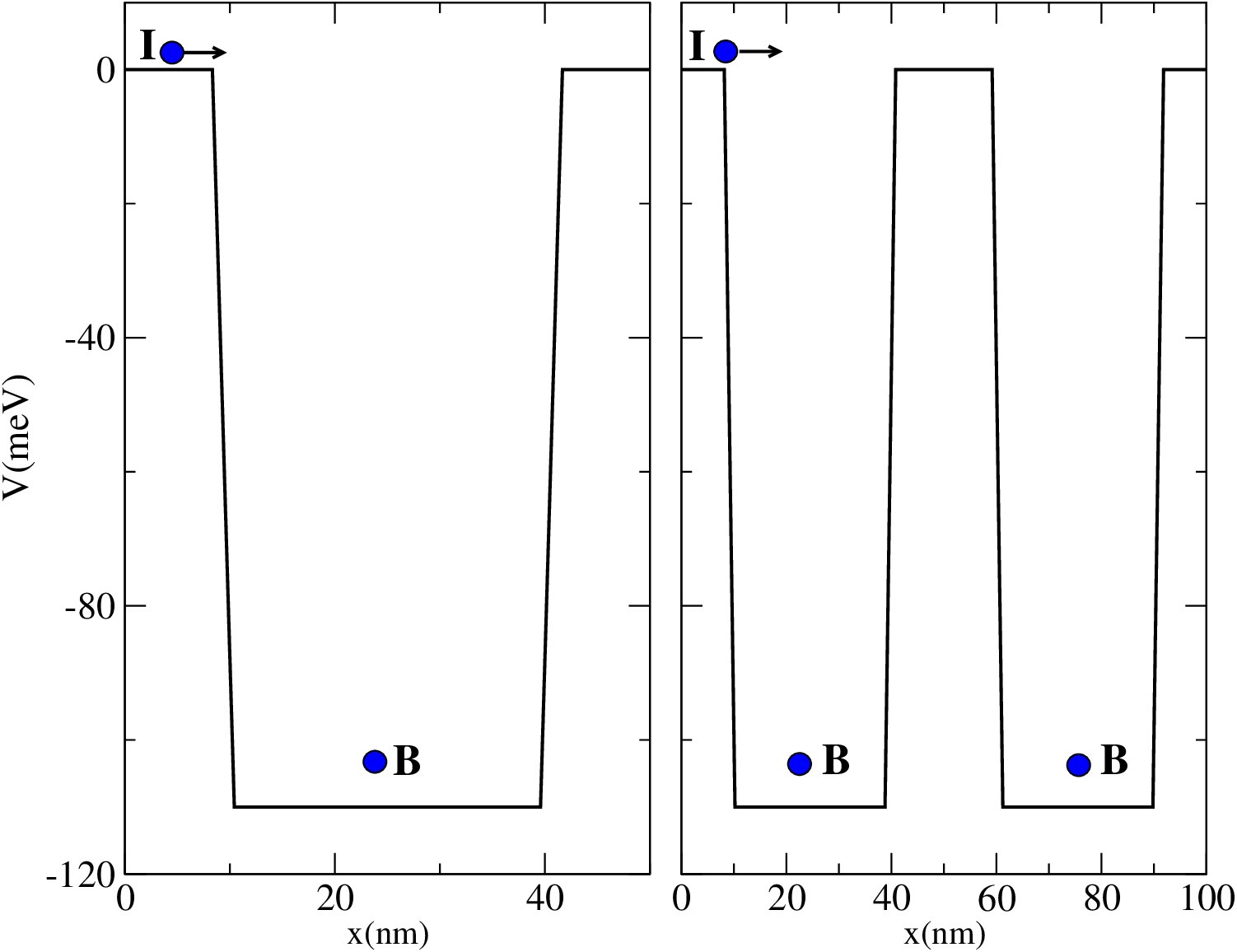}\vspace*{8pt}
    \caption{ \label{fig1} (Left panel) Potential profile $V_{QD}(x)$
      of the QD structure: the potential well is 110 meV deep and 30
      nm wide.   (Right panel) Potential profile $V_{DQD}(x)$ of the
      double-dots structure: the two potential wells are each 110 meV
      deep and 30 nm wide and are separated by a 20 nm long barrier.In both cases the
       incident electron (I) arrives from the left lead while the
      bound electron(s) (B) initialized in the (D)QD ground
      state.}
  \end{center}
\end{figure}

In this case  we consider a quasi one dimensional structure  where a potential
profile $V_{QD}$ mimicking a QD  is present.  The system is
sketched in the left panel of Fig.~\ref{fig1}.
The $N$ bound states and energies of the QD will be
indicated as $|\Phi_n\rangle$ and $E_N$ respectively, while $|T_{m}^{>(<)}\rangle$
describes the transmitted (reflected) electron with kinetic energy
$T_{m}$.  A single electron is localized in the QD ground state
$|\Phi_0\rangle$ and interacts, via Coulomb potential, with  a second
electron incoming from the left lead with kinetic energy $T_0$.
Specifically, we consider only cases in which $T_0$ is not sufficiently
high to ionize the QD. Moreover the charging energy of the latter
is assumed to be larger than the spacing between the the
single-particle energy levels. This means that our system always
operates in the two-particle regime.

The two-particle Hamiltonian $\mathcal{H}(x_1,x_2)$ is  given by
\begin{equation} \label{Ham}
  \mathcal{H}(x_1,x_2)=-\frac{\hbar^2}{2m^{\ast}}
  \frac{\partial^2}{\partial x_1^2}-\frac{\hbar^2}{2m^{\ast}}
  \frac{\partial^2}{\partial x_2^2}+V_{QD}(x_1) +V_{QD}(x_2)+
  \frac{e^2 }{4\pi \varepsilon r_{ij}},
\end{equation}
where $r_{ij}=\sqrt{\left(x_i-x_j\right)^2+d^2}$, $\varepsilon $ and
$m^{\ast}$ indicate the dielectric constant and effective mass of the
GaAs, respectively and $d$ is the cut-off term corresponding roughly
to the lateral dimension of the confinement\cite{Fogler}.

The two-particle scattering state $|\Psi\rangle$ is obtained by
solving the time-independent Schr\"odinger equation $\mathcal{H}
\Psi(x_1,x_2)= E \Psi(x_1,x_2)$ in the square domain $x_1,x_2\in[0,L]$
with $\mathcal{H}$ of Eq.~(\ref{Ham}).~\cite{berto2} To this aim a generalization of
the quantum transmitting boundary method\cite{Bertoni,Fogler,Lent} is used. When the scattered particle leaves the simulation domain, i.e. after the scattering took place, the two-particle wavefunction reads
\begin{eqnarray}\label{phiout}
  |\Psi_{QD}\rangle=\sum_{n=0}^M b_{n}|T_{n}^< \phi_n \rangle+\sum_{n=0}^M c_{n} |T_{n}^> \phi_n \rangle,
\end{eqnarray}
where the coefficients $b_{n}$ and $c_{n}$ represent the reflection
and transmission amplitudes in the different energy levels, respectively, and
the allowed energies $T_{n}$ of the scattered electron satisfy the
energy conservation $E_0 +T_0=E_n+T_n$. $M$ is the number of states
for which $T_n$ is positive: for our choice of simulation parameters  the contribution of the evanescent modes
to the output state is found negligible.  
  Here, all the travelling
components of the scattered carrier, both reflected and
transmitted, are taken into account in order to evaluate the appearance
of quantum correlations and therefore no post-selection process is
used, unlike other approaches.\cite{Lopez,Buscemi,busce2}

The calculation of the entanglement  between the
single-particle states of the QD and  the scattered particle
can be performed by means of the von Neumann entropy
$\xi_{QD}$ of the one-particle reduced density matrix $\rho_{QD}$
describing the bound electron.~\cite{Peres}  By tracing the two-particle density
matrix $\rho= |\Psi\rangle \langle \Psi|$ over the degrees of freedom
$|T_{n}^{>(<)}\rangle$ of the scattered electron, we obtain
$\rho_{QD}$ which takes a diagonal form. Its elements are given by
${\rho_{QD}}_{nn}$=$|b_n|^2+|c_n|^2$. Thus the von Neumann entropy can
be expressed as:
\begin{equation}
  \xi_{QD}=-\textrm{Tr}\left[\rho_{QD} \ln{\rho_{QD}}\right]=-\sum_{n=0}^M \left( |b_n|^2+|c_n|^2\right) \ln{\left( |b_n|^2+|c_n|^2\right)},
\end{equation}
and is bound in the interval $[0, \ln{(M+1)}]$, with $\xi_{QD}$=0
indicating no quantum correlations and $\xi_{QD}$=$\ln{(M+1)}$
indicating the maximum entanglement.

\subsection{Double-dot scattering}
In this subsection we describe the DQD system with three particles.
Here, a  Coulomb scattering
between a carrier propagating in a quasi 1D GaAs  channel with
energy $T_0$ and two electrons in a bound state of a potential
structure $V_{DQD}$ occurs. The potential  $V_{DQD}$ consists of  two wells separated by a barrier and
therefore mimicking a DQD device (see the right panel of 
Fig.~\ref{fig1}). The two-particle bound states of the latter will be
indicated as $|\Gamma_n\rangle$ and their energies as $\epsilon_N$.
Unlike the single QD, some DQD energy levels can be degenerate.  As
initial condition, we take the two electrons in the ground state
$|\Gamma_0\rangle$.

The three-particle Hamiltonian $\mathcal{H}(x_1,x_2,x_3)$ reads
\begin{equation} \label{Ham2}
  \mathcal{H}(x_1,x_2,x_3)=\mathcal{H}_0(x_1)+\mathcal{H}_0(x_2)+\mathcal{H}_0(x_3)+\sum_{i=1}^3
  \sum_{j=1}^{i-1}\frac{e^2 }{4\pi \varepsilon r_{ij}},
\end{equation}
where $\mathcal{H}_0(x_i)$ denotes the single-particle Hamiltonian
\begin{equation}
  \mathcal{H}_0(x_i)=-\frac{\hbar^2}{2m^{\ast}} \frac{\partial^2}{\partial x_i^2} + V_{DQD}(x_i).
\end{equation}
Also in this case, GaAs material parameters have been used
in  numerical calculations. By applying the  few-particle version of the 
\emph{quantum transmitting boundary method}
to solve the Schr\"odinger equation with
$\mathcal{H}(x_1,x_2,x_3)$, we find the reflection and the transmission
amplitudes $\beta_{n}$ and $\alpha_{n}$, in analogy with the previous case,
for the three-particle wavefunction:
\begin{eqnarray}\label{phiout2}
  |\Psi_{DQD}\rangle=\sum_{n=0}^M \beta_{n}|T_{n}^< \Gamma_n \rangle+\sum_{n=0}^M \alpha_{n} |T_{n}^> \Gamma_n \rangle.
\end{eqnarray}
  Also in
this case, the entanglement  between the scattered carrier and the  DQD system  can be
evaluated in terms of the von Neumann entropy $\xi_{DQD}$ of
the reduced density operator $\rho_{DQD}$ describing the two bound electrons.
 Due to the degeneracy of
some DQD energy levels, $\rho_{DQD}$ is a diagonal-block matrix with
the dimension of each block equal to the degree of degeneracy of the
corresponding energy level and thus its diagonalization is required to
calculate $\xi_{DQD}$  in terms of the eigenvalues
$\lambda_n$ as $\xi_{DQD}=-\sum_{n=0}^M \lambda_n
\ln{\lambda_n}$.

\section{Numerical Results}
Here we present the numerical results and  we analyze the
relation between resonances in transmission and reflection spectra and
the  quantum correlations.

\begin{figure}[htbp]
  \begin{center}
\includegraphics*[width=0.8\textwidth]{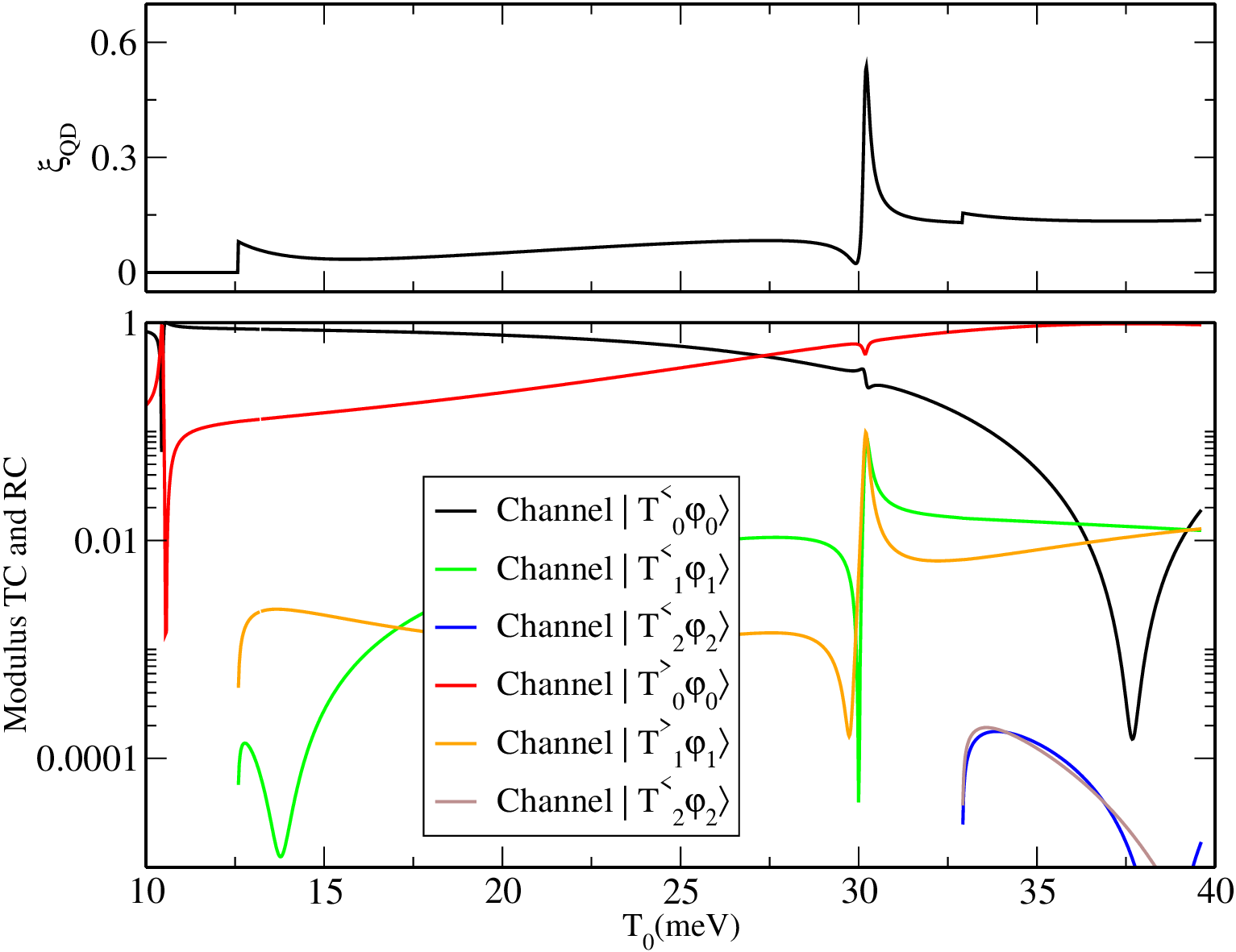}
 \caption{\label{fig2} (Top) Von Neumann entropy $\xi_{DQ}$
     and (bottom) modulus of the TCs and RCs 
    of the the  channels $T_0 E_0$,  $T_1 E_1$, and  $T_2 E_2$,
    as a function of the initial energy of the incoming electron $T_0$ 
    ranging from 10 to 40 meV,  for the two-particle scattering in the single QD nanostructure.}
\end{center}
\end{figure}

In the upper panel of  Fig.~\ref{fig2} we report  the von Neumann entropy $\xi_{QD}$ as a function of the initial energy of the incoming carrier for the
single-dot case, while in the lower panel  the
modulus of the RCs and TCs. 

\begin{figure}[htbp]
  \begin{center}
\includegraphics*[width=0.8\textwidth]{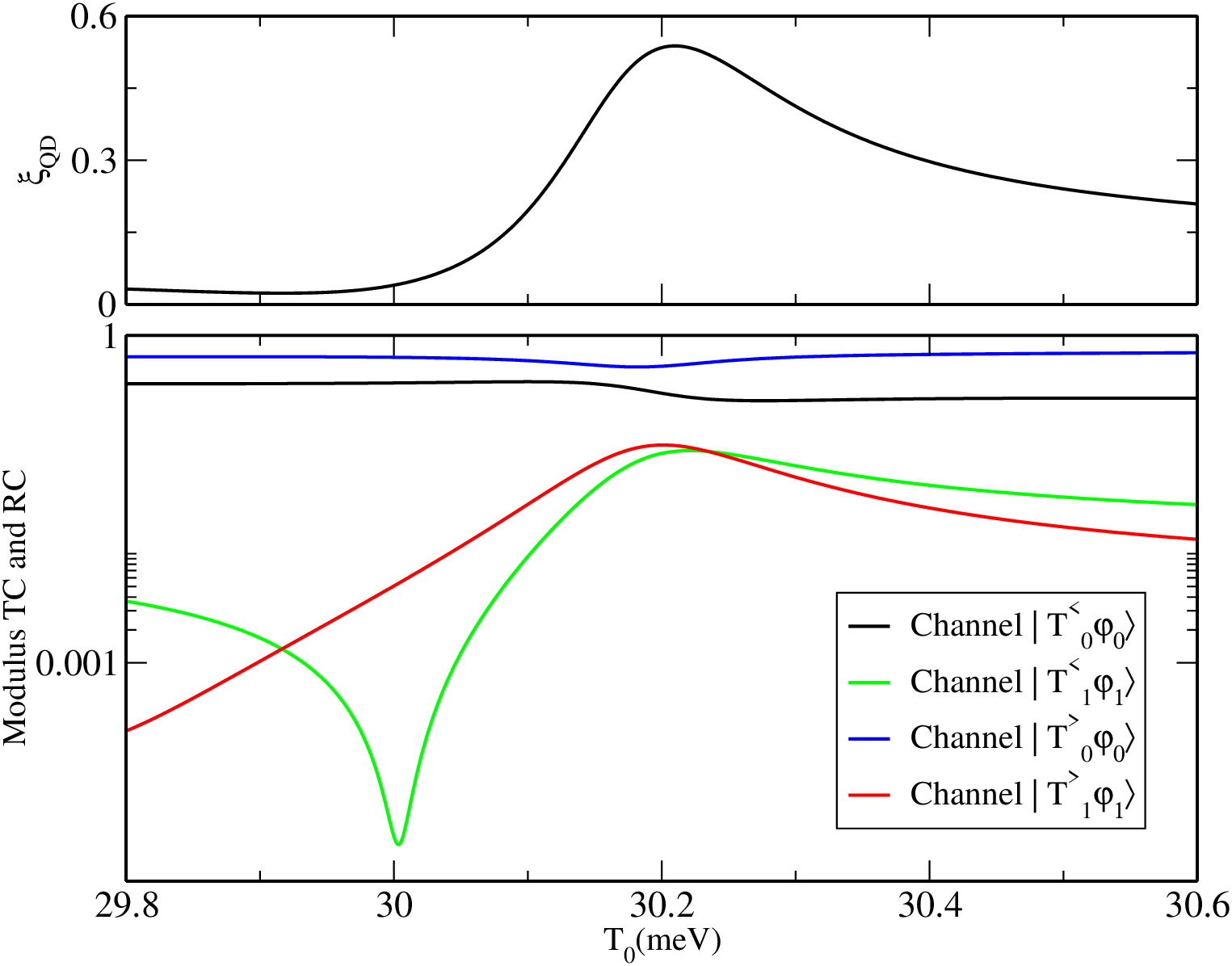}
 \caption{\label{fig2bis} (Top) $\xi_{DQ}$
      and (bottom) modulus of the TCs and RCs 
     of the   channels  $T_0 E_0$,  $T_1 E_1$  versus  $T_0$, close to a resonant     
   condition  for the two-particle scattering of  Fig.~\ref{fig2}.}
\end{center}
\end{figure}

At low energy no correlations between the single-particle
energy levels are found. In this case only a single channel
is present  for transmission and reflection: the incident carrier
is scattered elastically (either transmitted or reflected) with energy $T_0$ while the QD is left
in its ground state $\phi_0$. When the kinetic
energy of the incoming carrier reaches a threshold value, around 12.5 meV, a 
new channel comes into play.  Now, as a consequence of the  scattering,
the dot can also  be excited in the first level $\phi_1$ with the energy 
of scatterer decreased to  $T_1$. Starting from this energy, quantum correlations are built up between  the particles
and their peculiar behavior  can be related to  the resonances  appearing in transmission
and reflection spectra.\cite{Buscemi,busce2} In order to get a better insight into such
phenomena, a zoom of the above curves  around a resonant energy $T_0$=30 meV is displayed
in  Fig.~\ref{fig2bis}.  
 The entanglement  exhibits a maximum (where
$\xi_{DQ}$ is around 0.6),  
when both the RC and TC of the second channel exhibit a
peak. This means that the excitation probability  of the DQ due to the collision 
becomes comparable to the probability of an elastic scattering,
and this results in a significative increase in the lack
of knowledge about the state of the electron bound in the dot.

\begin{figure}[htbp]
  \begin{center}
\includegraphics*[width=0.8\textwidth]{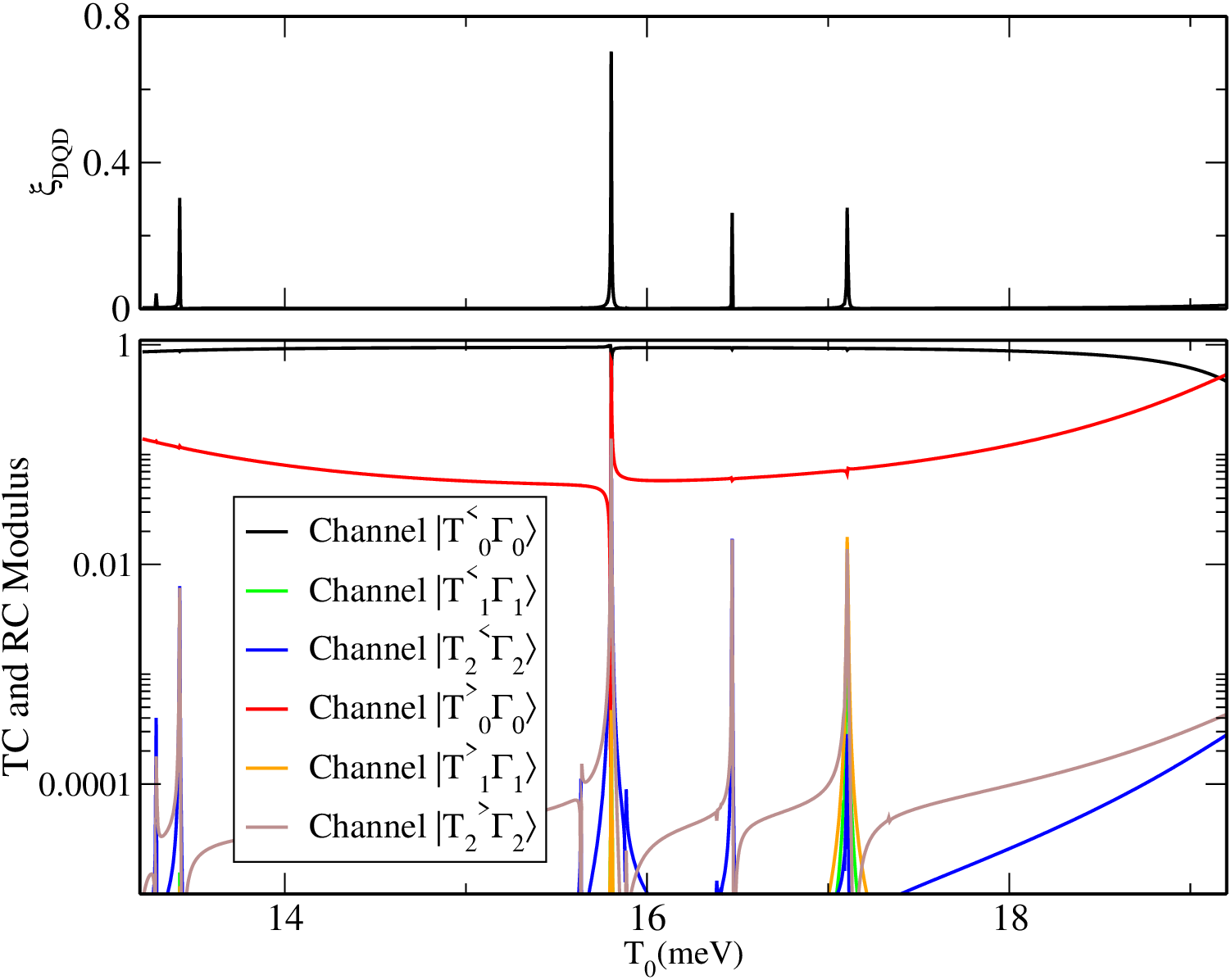}
 \caption{\label{fig3}  (Top) Von Neumann entropy $\xi_{DQD}$ and (bottom) modulus of the TCs and RCs 
    of the   channels $T_0 \epsilon_0$,  $T_1 \epsilon_1$, and  $T_2 \epsilon_2$,
    as a function of the initial energy of the incoming electron $T_0$ 
    ranging from 10 to 40 meV,  for the three-particle scattering in the DQD nanostructure.}
\end{center}
\end{figure}

\begin{figure}[htbp]
  \begin{center}
\includegraphics*[width=0.8\textwidth]{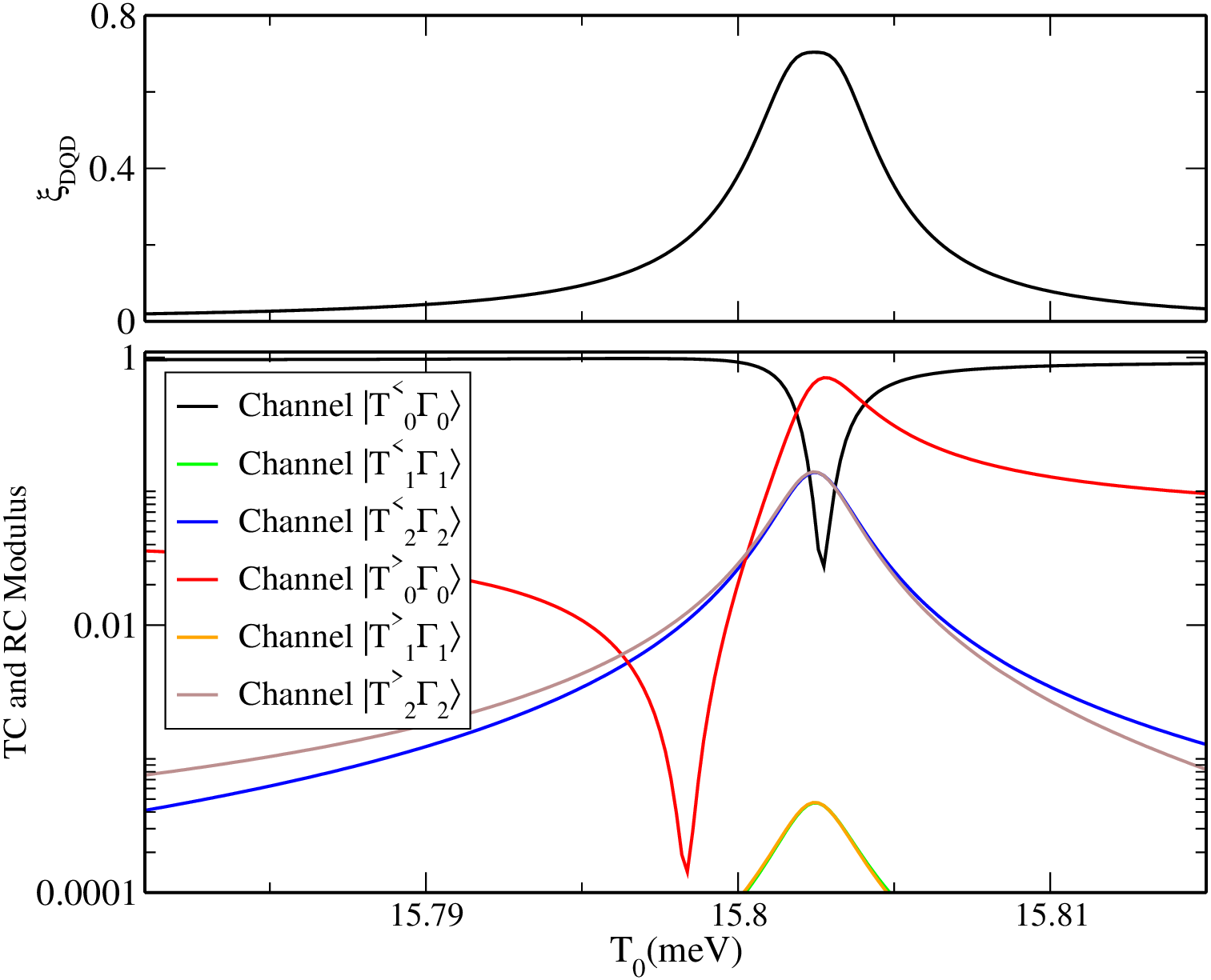}
 \caption{\label{fig3bis}  (Top)  $\xi_{DQD}$
      and  (bottom) modulus of the TCs and RCs 
     of the   channels  $T_0 E_0$,  $T_1 E_1$ versus  $T_0$, close to a resonant condition for the three-particle scattering of Fig.~\ref{fig3}.}
\end{center}
\end{figure}

Also the quantum correlations appearing in the DQD structure,
as a consequence of the  three-particle scattering event, can be related
to the resonances  in reflection and transmission spectra,
as can be seen from  Fig.~\ref{fig3}. 
The maximum
value of $\xi_{DQD}$ is reached when the initial energy $T_0$
of the incoming carrier is around 15.8 meV (see the detail expanded in  Fig.~\ref{fig3bis}). 
This peak occurs when the sum of  the moduli of TC and  RC
describing the elastic scattering, i.e. when the DQD is left in its ground state $\Gamma_0$
with the scattered carrier in the energy level $T_0$, attains its minimum value
while the probability that the scattering occurs through other channels
is maximal. Therefore  the energy levels of the DQD particles
are  correlated to the output energies of the scattered electron
this meaning  that a single pure  state cannot be ascribed to the DQD system
after the scattering.
\section{Conclusions}
In this paper we have calculated by means of 
the von Neumann entropy the amount of  quantum correlations
due to two- and three-electron scatterings  in QD and DQD
nanostructures, respectively. The outcomes of the 
 numerical simulations  show that the 
 quantum entanglement between the scattered carrier  and  the particle(s) bound in the single
or double dot systems is strictly related to the behavior
of the TC and RC of the  various scattering channels.\cite{Lopez,Buscemi,busce2,Kazu} 
Specifically, when  the scattering occurs through various  channels
with comparable probabilities, the uncertainty about the  specific
energy states of the electrons bound in QD and DQD is increased   due
to the greater amount of entanglement  with the scattered particle.

Our results  agree
with previous works\cite{Lopez,Buscemi,busce2} where the entanglement
formation due to the scattering events  in single QDs
is shown to be  related to the different nature of
the resonances present only in the transmission spectra of the 
injected carrier. There the entanglement is an estimation
of the  correlations showed up once the crossing through the QD
is successfully detected. On the other hand, in this work  all the travelling
components,  either reflected or transmitted, of the
scattered electron have been taken into account,
that is no post-selection process has been considered. 
Thus our investigation  
can be viewed as the first step 
towards a rigorous   analysis  of the 
decoherence   due to successive scatterings  of a number
of  injected carriers, and therefore of an electric  current.

\section*{Acknowledgments}
We are pleased to thank Carlo Jacoboni for fruitful discussions.
We  acknowledge support from CNR
Progetto Supercalcolo 2008 CINECA.

\end{document}